\documentclass[12pt]{iopart}

\usepackage{epsfig}

\begin{document}
\title[Critical properties of the susceptible-exposed-infected
model on a square lattice]
{Critical properties of the susceptible-exposed-infected model
on a square lattice}

\author{Alexander H. O. Wada, T\^ania Tom\'e and M\'{a}rio J. de Oliveira}

\address{Instituto de F\'{\i}sica,
Universidade de S\~{a}o Paulo,
Caixa Postal 66318
05314-970 S\~{a}o Paulo, S\~{a}o Paulo, Brazil}

\ead{oliveira@if.usp.br}

\begin{abstract}

The critical properties of the stochastic 
susceptible-exposed-infected model on a square lattice
is studied by numerical simulations and by the use of
scaling relations. In the presence of
an infected individual, a susceptible becomes either 
infected or exposed. Once infected or exposed, the
individual remains forever in this state. 
The stationary properties are shown to be the same as those of 
isotropic percolation so that the critical behavior
puts the model into the universality class of
dynamic percolation.

\end{abstract}


\maketitle

\section{Introduction}

Spatio-temporal structures as well as fluctuations are essential
features of epidemic spreading \cite{bailey57,nisbet82,renshaw91,mollison95}. 
A description of epidemic spreading that takes into account these
essential features is provided by stochastic lattice models
\cite{satulov94,durrett95,marro99,
keeling99,antal01a,dammer03,arashiro07,henkel08,souza10,souza11,tome11,souza13}
in which each site of a lattice is occupied by an individual that
can be in one of a certain number of states.
In the susceptible-infected-recovered (SIR) model 
\cite{souza10,souza11,tome11},
an important model in this context,
the possible states are susceptible (S), infected (I) or recovered (R). 
The SIR model is composed by two  processes. One in which a
susceptible becomes infected by a catalytic reaction, S+I$\to$I+I,
and another in which an infected becomes recovered spontaneously,
I$\to$R.
Another model, the one that will be the object our study here, is the 
susceptible-exposed-infected (SEI), introduced by Tom\'e and de Oliveira
\cite{tome11}, in which each individual
can be susceptible (S),  exposed (E) or infected (I). This model has also two
processes. In the presence of an infected individual,
a susceptible individual may become either infected or exposed,
processes represented by the reactions S+I$\to$I+I and S+I$\to$E+I,
respectively.

The distinguish features of these two models are as follows. 
The spreading of the epidemic occurs as long as there are
active sites, which are the sites occupied by a susceptible individual
next to an infected individual.
When the active sites have disappeared, the infection reaction S+I$\to$I+I
no longer takes place, that is, no new infected individuals are created,
and the whole process eventually stops.
The system finds itself in one of many absorbing states
which, in the SIR model, are the configurations formed by R and S sites
and, in the SEI model, are the configurations without any pair
of neighboring SI sites. 
Starting from a configuration full of susceptible individuals except
for a single infected site, the process generates a cluster of 
inactive sites, which are the R sites of the SIR model and the I sites
of the SEI model. In the stationary state, which is an 
absorbing state, these clusters have the same properties
as the clusters occurring in isotropic
percolation model so that the stationary properties
of these two models are similar to percolation. 

When the rate of infection is small there is no epidemic
spreading, that is, the generated clusters are all finite.
If however the infection rate is large enough, an infinite
cluster of inactive sites is generated and
the epidemic spreading takes place. The transition
from non-spreading to spreading is regarded as a
continuous phase transition with critical behavior
within the universality class of dynamical percolation 
\cite{antal01a,dammer03,arashiro07,henkel08,souza10,souza11,
grassb83,cardy83,cardy85,munoz99}.
In fact, the clusters generated by the rules of the SEI model
can be exactly mapped into the clusters of site percolation
so that the stationary properties of the SEI model are identical
to the properties of site percolation. In this sense,
it is similar to the model introduced by Alexandrowicz
\cite {alexandrowicz80} to generate percolation clusters.
The SEI model can thus be regarded as a standard
example of a model belonging
to the dynamical percolation universality class.
Here we are concerned with the critical behavior of
the SEI model on a square lattice, particularly with
the numerical calculation of the dynamic critical exponents.
As to the static critical exponents, they are the
same as the percolation in two dimensions and known
exactly \cite{stauffer85}. 
%
%
Here we perform calculation for the SEI model with
results for the dynamic critical exponents that are very accurate, and
in agreement with the exponents of the dynamic percolation
universality class \cite{souza10,munoz99}.

\section{Model}

The SEI model is a continuous time stochastic markovian 
process defined on a lattice where each site can be in
one of three states: S, I or E. The allowed transitions
are those in which just one site changes its state.
The transition rate of the process S$\to$I is $b f$,
where $f$ is the fraction of I sites in the neighborhood
of the site to be updated and $b$ is the infection parameter.
The transition rate of the process S$\to$E is $a f$,
where $a$ is the exposition parameter.
Other transitions are forbidden so that sites in states
I or E remain forever in these states.
For convenience, we define the parameters $p=b/(a+b)$
and $q=a/(a+b)$ so that $p+q=1$. 
The model displays two regimes. One in which there is no
epidemic spreading, occurring for small infection rate,
and the other in which the epidemic spreading takes place,
occurring for large infection rate.
The stationary properties are close related to site percolation
model. In fact, the cluster of I sites can be exactly mapped into
the cluster of site percolation in the same lattice, in 
which each site is occupied with probability $p$,
as will be shown below. It is well known that 
the percolation model \cite{stauffer85} shows a phase transition
from a state with finite clusters, occurring for $p<p_c$, to a state 
with an infinite cluster, or percolating state, for $p>p_c$, 
where $p_c$ is the critical concentration. We thus expect
for the SEI model, a transition from a non-spreading
regime, for $p<p_c$, to a spreading regime, for $p>p_c$.

The simulation of the model is carried out as follows.
Each site of a regular lattice
with $N$ sites can be in one of three states: occupied by a
susceptible (S), by an infected (I) or by an exposed (E) individual.
At each time step a site is chosen at random. If it is in
the I or E states nothing happens. If it is in the state S then
with probability $pf$ it becomes I and with probability $qf$ it becomes E,
where $f$ is the fraction of I sites in the neighborhood of the chosen site
and $p+q=1$. Equivalently, if the chosen site is in state S,
we may randomly choose a neighboring site;
if it is in the I state then the chosen site becomes
I with probability $p$ or E with the complementary probability
$q=1-p$. The time is increased by an amount equal to $1/N$.
The neighborhood of a site is defined as its nearest neighbor sites.

An alternative approach, useful for time dependent simulations,
is carried out as follows. At each time step, a site 
is chosen at random from a list comprising the I sites only.
Next, one of its neighbors is chosen at random. If this neighbor
is in state I or E nothing happens. If it is in state S, then
it changes to I with probability $p$ or to E with probability
$q=1-p$. The time is increased by an amount equal to $1/n_I$
where $n_I$ is the number of sites in the list, which is 
the total number of I sites in the lattice, and the list is updated.
An even more efficient algorithm is set up by using a list
of active pair of sites, more precisely, a list of nearest neighbor pairs of
susceptible-infected sites. At each time step, a pair of the list
is chosen at random and the S site of the pair becomes
I with probability $p$ or E with probability $q=1-p$. 
The time is increased by an amount $1/n_{SI}$ where $n_{SI}$
is the number of entries in the list, which is 
the total number of SI pairs in the lattice, and the list is updated.

Let us consider a finite lattice full of susceptible individual
except for one site which is occupied by an infected individual.
The system evolves in time and eventually reaches its final
state which comprises a connected cluster of infected sites
in addition to exposed and susceptible sites.
The exposed sites are found at the boundary of the cluster
separating the infected sites from the susceptible sites,
so that active sites are absent. The clusters generated by
the SEI rules are the same clusters of site percolation,
as we show next, in which
the sites are occupied independently with probability $p$.

The mapping of the stationary properties of the SEI model into the
site percolation model can be understood as follows. 
Suppose that a cluster C of infected sites of a lattice
has been generated by one of the algorithm above, starting from an 
infected site at the origin. Suppose moreover that
the dynamics has come to a halt so that there are no pairs of the type
SI. The boundary B of this cluster is therefore composed only by E sites.
During the dynamics, whenever the site $i$ becomes either
I or E, we keep the used random number $\xi_i$.
It is clear that, if site $i$ has turned into an I site then
$\xi_i\leq p$; if site $i$ has turned into an E site then $\xi_i> p$.
Let us now consider a replica of the lattice, with all sites empty except the
site at the origin which is occupied. Next the site $i$ of the replica
is occupied if $\xi_i\leq p$ and remains empty if $\xi_i> p$.
By means of this procedure, which is the procedure used in site percolation, 
a cluster of occupied sites is generated, which is thus
identical to the cluster of I sites of the original lattice.

\section{Scaling relations}

Around the critical point, the quantities that
characterize the system are assumed to obey a scaling relation. 
We assume two types of scaling relation. The first one
is a finite-size scaling relation, valid at the stationary state.
A certain quantity $Q$ depends
on the linear size of the system $L$ and on the
deviation $\varepsilon=p-p_c$ according to
\begin{equation}
Q(\varepsilon,L) = L^{x/\nu_\perp}\Phi(\varepsilon L^{1/\nu_\perp}),
\label{1}
\end{equation}
where $\nu_\perp$
is the critical exponents related to the spatial correlation length
and $x$ is the critical exponent related to $Q$ in the
thermodynamic limit, that is, $Q\sim \varepsilon^{-x}$ when $L\to\infty$.
The second type is a time dependent scaling relation,
valid for an infinite system, and given by
\begin{equation}
Q(\varepsilon,t) = t^{\,x/\nu_\parallel}
\Psi(\varepsilon t^{\,1/\nu_\parallel}),
\label{2}
\end{equation}
where $\nu_\parallel$ is the critical exponents related to
the time correlation length. When $t\to\infty$, we get the same
behavior $Q\sim \varepsilon^{-x}$.
At the critical point, relations (\ref{1}) and (\ref{2}) 
predict the following scaling forms
\begin{equation}
Q\sim L^{x/\nu_\perp}, \qquad\qquad
Q\sim t^{y},
\end{equation}
respectively, where $y=x/\nu_\parallel$.

The behavior of the SEI model 
is characterized by a set of quantities.
We define the average  $N_I=\langle n_I \rangle$
where $n_I$ is the number of infected sites
and the average $N_{SI}=\langle n_{SI}\rangle$ in the number
of active pair of sites defined as the number $n_{SI}$
of pairs of neighboring sites of type SI.
Another relevant quantity
is the surviving probability $P(t)$, defined as the 
probability that at time $t$ the system is active,
that is, as long $n_{SI}\neq 0$.
Starting from one infected site in a infinite lattice full of susceptible,
we expect the  following asymptotic time behavior, at the critical point,
\begin{equation}
N_I(t) \sim t^{\,\eta},
\label{5}
\end{equation}
\begin{equation}
N_{SI}(t) \sim t^{\,\theta},
\label{6}
\end{equation}
\begin{equation}
P(t) \sim t^{-\delta},
\label{7}
\end{equation}
\begin{equation}
\xi(t) \sim t^{1/z}.
\label{8}
\end{equation}
where $\xi$ is the spatial correlation length and
$z=\nu_\parallel/\nu_\perp$.

If several trials are carried out, some survive up to time $t$,
some do not. In the limit $t\to\infty$, the cluster of a surving
trial will be identified with the infinite percolating
cluster. Let $n_s$ be the variable that counts
the number of infected sites in the surviving trials.
Denoting by $M$ and $M_s$ the number of trials and the number
of surviving trials, respectively, and by $n_{si}$ the number of infected sites 
in the $i$-th surviving trial, then the average $N_s$ of $n_s$ 
over the surving trials is given by 
\begin{equation}
N_s = \frac{1}{M_s}\sum_i n_{si} 
= \frac{M}{M_s}\left(\frac{1}{M}\sum_i n_{si}\right).
\end{equation}
The quantity between parentheses, which we denote by $N_I^\prime$,
is the average number of $n_s$ over all trials because $n_s=0$
for a nonsurviving trial. Taking into account
that $P=M_s/M$ then $N_s=N_I^\prime/P$.
Now, the quantity $N_I^\prime$ has a behavior similar to
the average $N_I$ of $n_I$ over all trials. In fact, we found
numerically that $N_I^\prime$ is proportional to $N_I$ for large enough $t$.
Therefore, for large enough $t$ we may write $N_s\sim N_I/P$.
Taking into account relations (\ref{5}) and (\ref{7}),
\begin{equation}
N_s(t) \sim t^{\,\eta+\delta}.
\label{9}
\end{equation}

The number of infected sites in surviving trials 
inside a region of linear size $\xi$ is proportional to the order
parameter $P$. This amounts to say that the ratio 
$N_s/\xi^d$ is proportional to $P$.
Writing thus
$N_s/\xi^d\sim P$, and taking into account relations (\ref{7}), (\ref{8})
and (\ref{9}), we reach the following exponent relation
\begin{equation} 
z(\eta +2\delta) = d.
\label{13a}
\end{equation}

At the critical point a surviving trial, which makes up the
percolating cluster, has a fractal structure with a fractal
dimension $d_F$ defined by 
\begin{equation}
N_s \sim \xi^{\,d_F}.
\label{23}
\end{equation}
Comparing (\ref{8}) and (\ref{9}) with (\ref{23}), we see that
the fractal dimension $d_F$ of the critical cluster is related
to $\eta$ and $\delta$ by
\begin{equation}
z(\eta +\delta) = d_F.
\label{14}
\end{equation}

The exponents $\theta$ and $\eta$ are connected by the relation
\begin{equation}
\eta = 1+\theta,
\label{10}
\end{equation}
which can be understood by observing that the rate of increase in the 
average number of infected sites is proportional to the number
of active pair of sites, that is, 
\begin{equation}
\frac{d}{dt}N_I = \frac{b}{k}N_{SI}
\label{12}
\end{equation}
where $k$ is the number of neighbors.
Replacing (\ref{6}) into (\ref{12}), it follows that $N_I\sim t^{1+\theta}$
from which we find relation (\ref{10}).
We should remark that in models belonging to
direct percolation (DP) universality class \cite{henkel08}, 
$N_I\sim N_{SI}$
so that the exponents $\eta$ and $\theta$ coincide.
In the case of models belonging to dynamical percolation
universality class, such as the SIR and SEI models,
they are distinct and are related by (\ref{10}).

\begin{figure}
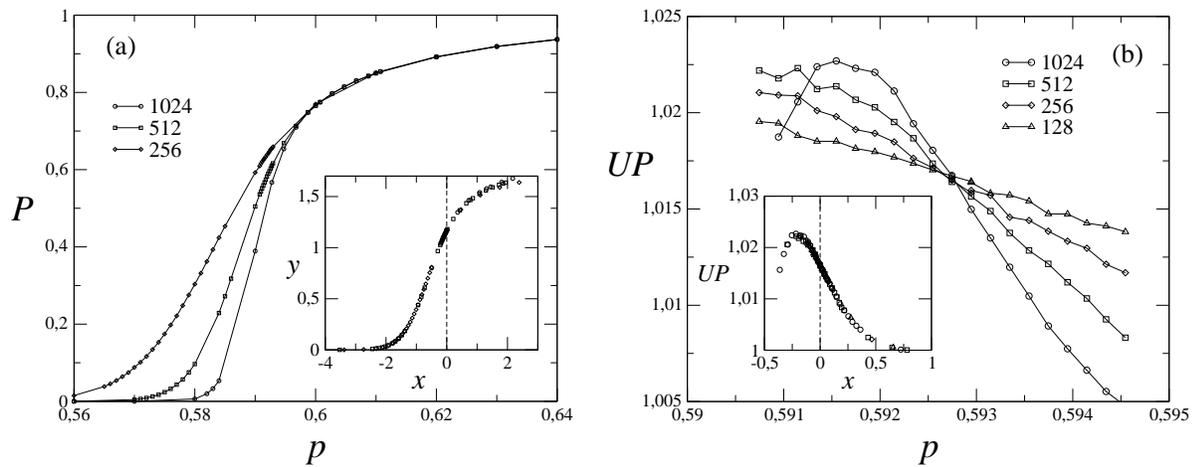

\centering
\epsfig{width=7.5cm,file=PXp.eps}
\hfill
\epsfig{width=7.8cm,file=UPXp.eps}
\caption{Static properties of the SEI model from numerical simulations
on a square lattice. (a) Order parameter $P$ versus $p$
for several values of $L$ indicated. The inset is a data collapse
showing $y=PL^{\beta/\nu_\perp}$ versus $x=\varepsilon L^{1/\nu_\perp}$
where $\varepsilon=p-p_c$. (b) The quantity $U\!P$ versus $p$ for several
values of $L$ indicated. The inset is a data collapse
showing $U\!P$ versus $x$.} 
\label{estac}
\end{figure}

When $t\to\infty$, that is, in the stationary state,
the surviving probability, which is identified with
the order parameter, behaves around the critical point as
\begin{equation}
P \sim \varepsilon^{\beta} 
\end{equation}
and the exponent $\beta$ is related to $\delta$ by 
\begin{equation}
\delta\nu_\parallel=\beta,
\label{15a}
\end{equation}
which is equivalent to $\delta z = \beta/\nu_\perp$.
Taking into account this relation and
comparing relations (\ref{14}) and (\ref{13a}), 
we find $\eta z = \gamma/\nu_\perp$, where
$\gamma=d\nu_\perp-2\beta$, and $(d-d_F)\nu_\perp=\beta$.

The density $\rho$ of infected sites is equal to
to $N_I/N$. Bearing in mind that $N_I=N_s P$ and that
$N_s=PN$, we conclude that $\rho\sim P^2$ so that
\begin{equation}
\rho \sim \varepsilon^{2\beta}.
\end{equation}
Let us define the quantity $\rho_2=\langle n_I^2\rangle/N^2$.
Taking into account that  $\langle n_s^2\rangle=\langle n_I^2\rangle/P$
and that $\langle n_s^2\rangle/N^2\sim P^2$ around the critical point,
we may conclude that $\rho_2\sim P^3$, or
\begin{equation}
\rho_2 \sim \varepsilon^{3\beta}.
\end{equation}
The quantity $U$, defined as
the ratio $U=\langle n_I^2 \rangle/\langle n_I \rangle^2=\rho_2/\rho^2$
behaves as 
\begin{equation}
U \sim \varepsilon^{-\beta}.
\end{equation}

Scaling relations can be written for the quantities defined above.
For instance, the  order parameter obeys the scaling relation
\begin{equation}
P(\varepsilon,L) = L^{-\beta/\nu_\perp}\Phi_1(\varepsilon L^{1/\nu_\perp}).
\label{15}
\end{equation}
From the above relations for $U$ and $P$, 
it follows that the product $U\!P$ approaches a constant at
the critical point and obeys the relation \cite{souza11}
\begin{equation}
U\!P = \Phi_2(\varepsilon^{\,\nu_\perp} L), 
\label{13}
\end{equation}
so that, at the critical point, $\varepsilon=0$, a plot of $U\!P$
versus $p$ for several values of $L$ will cross at the same point,
the critical point.

\begin{figure}
\centering
\epsfig{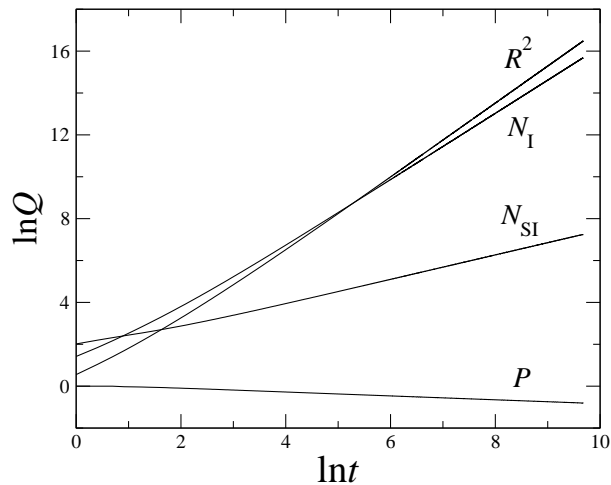}
\caption{ 
Number of infected sites $N_I$, number of active pair of sites $N_{SI}$,
surviving probability $P$ and spreading of the infected sites $R^2$
as a function of time $t$ in a double-log plot.
The slopes of the curves for large enough time give, respectively,
$\eta$, $\theta$, $-\delta$ and $2/z$. 
}
\label{qcm}
\end{figure}

\section{Simulations}

We performed numerical simulation on a square lattice
by using the algorithms explained above.
The stationary properties were obtained on
lattices of several sizes, and for several values of the parameter $p$. 
The order parameter $P$ was obtained as follows \cite{souza11}.
We perform
several runs starting from an infected individual
placed in the center of a finite lattice full of susceptible individuals. 
The quantity $P(p,L)$ is the fraction of runs such that 
an infected individual reaches the border of the lattice of linear size $L$.
In figure \ref{estac}a we show the order parameter $P$ versus $p$ 
for several values of $L$. 
The inset of figure \ref{estac}a shows a data colapse 
according to the scaling form (\ref{15}).
To get the data colapse we used the exact values of the percolation
exponents in two dimensions \cite{stauffer85}: $\beta=5/36$ and $\nu_\perp=4/3$.
We used also the numerical value of $p$ at the critical
point on a square lattice, $p_c=0.59274606(5)$ obtained from Monte Carlo
simulations \cite{feng08}.

In figure \ref{estac}b, we show the product $U\!P$ as a
function of $p$. As expected from scaling relation (\ref{13})
the curves for distinct $L$ cross at the critical point.
The value of $U\!P$ at $p_c$ is estimated to be
$(U\!P)_c=1.01658(1)$, which is a linear extrapolation
in $1/L^2$ obtained
from the values 1.016598, 1.016585, 1.016582
of this quantity for $L=256,512,1024$ respectively.
The quantity $U\!P$ at $p_c$ is a universal quantity
and may be compared to the values 1.0167(1) obtained
for percolation and SIR models \cite{souza11}.
The inset of figure \ref{estac}b shows the data collapse according to (\ref{13}).
Again we use the exact value $\nu_\perp=4/3$.

To get the dynamic exponents defined by relations (\ref{5}),
(\ref{6}), (\ref{7}), (\ref{8}), we performed time-dependent
Monte Carlo simulations at the critical point $p_c=0.59274606(5)$ \cite{feng08}.
To get the exponents $\eta$, $\theta$ and $\delta$, we have
estimated the number of infected sites $N_I$, the number of active pair of sites
$N_{SI}$ and the surviving probability $P$. The exponent $z$ was
obtained by calculation the spreading of the infected sites
$R^2$ \cite{marro99}, defined by 
\begin{equation}
R^2 = \frac{1}{N_I}\sum_i r_i^2 \langle\eta_i\rangle
\end{equation}
where $r_i$ is the distance of site $i$ to the origin
and $\eta_i$ takes the value 1 when site $i$ is occupied
by an infected individual and zero otherwise.
At the critical point
\begin{equation}
R^2 \sim t^{2/z}.
\end{equation}
Figure \ref{qcm} shows the quantities $N_I$, $N_{SI}$, $P$ and
$R^2$ as a function of time. Each curve corresponds to an
average over $10^6$ runs obtained on a square lattice of
linear size $L=2^{15}$. At each run, the lattice was full
of susceptible sites except the central site, which is in
an infected state. Up to the maximum time shown in figure \ref{qcm}
no infected site reached the border of the lattice, which
amounts to say that the results shown in figure \ref{qcm} are valid for 
an infinite lattice. We remark that the quantity $N_I^\prime$
was also calculated and we found $cN_I^\prime=N_I$ for large
enough $t$, as we have assume above, 
and that at the critical point $c=1.025(5)$.

\begin{table}
\begin{center}
\caption{\label{exponents} Critical exponents obtained by
fitting the scaling form (\ref{5}) to the curves
in figure \ref{qcm} within the time interval between $t_0$
and $t_1$. The last row gives the average of the values in
the previous rows, together with the errors in the last digit.}
\bigskip
\begin{tabular}{|l|l|l|l|l|l|}
\hline
$\ln t_0$ & $\ln t_1$ & $\eta$ & $\theta$ & $\delta$ & $z$  \\
\hline
\hline
$4.5$ & $9.5$ & $1.58462$   & $0.58444$   & $0.09217$   & $1.13096$   \\
$5.5$ & $9.5$ & $1.58457$   & $0.58438$   & $0.09208$   & $1.13087$  \\
$6.5$ & $9.5$ & $1.58454$   & $0.58442$   & $0.09196$   & $1.13083$  \\
\hline
      &       & $1.5846(2)$ & $0.5844(2)$ & $0.0921(3)$ & $1.1309(3)$ \\
\hline
\end{tabular}
\end{center}
\end{table}

To estimate the exponent $y$ for a given quantity $Q\sim t^y$,
we used a correction to scaling of the form \cite{marro99}
\begin{equation}
Q(t) = t^y(c_1+c_2 t^{-\mu}).
\label{25}
\end{equation}
The fitting of this form to the data points gives the exponent
$y$ and $\mu$ in addition to the constants $c_1$ and $c_2$.
The fitting was done within a certain interval of time
starting at time $t_0$ and ending at time $t_1$.
We have found that the best fittings give
$\mu$ around 1, but the actual values found for the exponent $y$
are not too sensitive to the value of the exponent $\mu$.
We used three values of $t_0$ and the same value of $t_1$.
As seen in table \ref{exponents},
the three values of a given exponent are distinct but very similar
allowing the estimation of the statistical error
shown in the last row together with the average.

The estimated exponents can be checked by using relations
$\eta z=\gamma/\nu_\perp$ and $\delta z=\beta/\nu_\perp$. 
From the values of $\eta$ and $z$ we get $\eta z=1.7920(7)$
which should be compared with the exact result 
$\gamma/\nu_\perp=43/24=1.791666\ldots$ \cite{stauffer85}.
From the values of $\delta$ and $z$ we get $\delta z=0.1042(4)$
which should be compared with the exact result 
$\beta/\nu_\perp=5/48=0.104166\ldots$ \cite{stauffer85}.
We remark that our results are in agreement with
results for the exponents of models belonging to
the universality class of dynamic percolation
\cite{souza10,munoz99}.
Notice that the critical exponents $\eta$ and $\theta$ satisfy, within the
statistical error, the relation (\ref{10}), $\eta=1+\theta$.

\section{Discussion}

We have determined the critical properties of the SEI model
on a square lattice by numerical simulations. The model is
an example of model belonging to the universality class
of dynamical percolation. The stationary properties are
shown to be exactly the same as those of isotropic percolation,
so that the static exponents are the same as percolation,
which in two dimensions are known exactly. This is not
the case of the dynamic exponents which we have calculated
here by using a time-dependent simulations. Our numerical
estimation, within the statistical errors, are in good
agreement with the known relations among the exponents.
An important feature that distinguish this model from
models belonging to the DP universality class \cite{henkel08}, such
as the susceptible-infected-removed-susceptible (SIRS) \cite{souza10},
rests on the relation between the number of infected sites and
the number of active pairs of sites. In the SIRS model these two 
quantities are closed related leading to the identification
of the exponents $\eta$ and $\theta$. In the SEI model
they are distinct leading to the relation $\eta=1+\theta$,
which was confirmed by our numerical simulations.





\section*{References}

\begin{thebibliography}{99}

\bibitem{bailey57} N. T. J. Bailey, {\it The Mathematical Theory
of Epidemics} (Hafner, New York, 1957).

\bibitem{nisbet82} R. M. Nisbet, Modeling Fluctuation Populations,
Wiley, New York, 1982.

\bibitem{renshaw91} E. Renshaw, {\it Modelling Biological Populations in
Space and Time} (Cambridge University Press, Cambridge, 1991).

\bibitem{mollison95} D. Mollison (ed.), {\it Epidemic Models}
(Cambridge University Press, Cambridge, 1995).

\bibitem{satulov94} J. Satulovsky and T. Tom\'e, Phys. Rev. E
{\bf 49}, 5073 (1994). 

\bibitem{durrett95} R. Durrett, ``Spatial epidemic models'', 
in \cite{mollison95}, p. 187. 

\bibitem{marro99} J. Marro and R. Dickman, {\it Noequilibrium
Phase Transitions in Lattice Models}, (Cambridge University Press,
Cambridge, 1999).

\bibitem{keeling99} M. J. Keeling, 
Proc. R. Soc. Lond. B {\bf 266}, 859 (1999).

\bibitem{antal01a} T. Antal, M. Droz, A. Lipowski and G. \'{O}dor, 
Phys. Rev. E {\bf 64}, 036118 (2001).

\bibitem{dammer03} S. M. Dammer and H. Hinrichsen, 
Phys. Rev. E {\bf 68}, 016114 (2003).

\bibitem{arashiro07} E. Arashiro and T. Tom\'e, 
J. Phys. A {\bf 40}, 887 (2007).

\bibitem{henkel08} M. Henkel, H. Hinrichesen and S. L\"ubeck,
{\it Non-Equilibrium Phase Transitions}, Vol. I:
{\it Absorbing Phase Transitions} (Springer, Dordrecht, 2008).

\bibitem{souza10} D. R. Souza and T. Tom\'e, 
Physica A {\bf 389}, 1142 (2010).

\bibitem{souza11} D. R. Souza, T. Tom\'e and R. M. Ziff,
J. Stat. Mech. P03006 (2011).

\bibitem{tome11} T. Tom\'e and M. J. de Oliveira,
J. Phys. A: Math. Theor. {\bf 44}, 155001 (2011).

\bibitem{souza13} D. R. de Souza, T. Tom\'e, S. T. R. Pinho, F. R. Barreto
and M. J. de Oliveira, Phys. Rev. E {\bf 87}, 012709 (2013).

\bibitem{grassb83} P. Grassberger, Math. Biosci. {\bf 63}, 157 (1983).

\bibitem{cardy83} J. L. Cardy, J. Phys. A {\bf 16}, L709 (1983)

\bibitem{cardy85} J. L. Cardy and P. Grassberger, 
J. Phys. A {\bf 18}, L267 (1985).

\bibitem{munoz99} M. A. Mu\~noz, R. Dickman, A. Vespignani,
and S. Zapperi, Phys. Rev. E {\bf 59}, 6175 (1999).

\bibitem{alexandrowicz80} Z. Alexandrowicz,
Phys. Lett. A {\bf 80}, 284 (1980).


\bibitem{stauffer85} D. Stauffer, {\it Introduction to Percolation Theory}
(Taylor and Francis, London, 1985).

\bibitem{feng08} X. Feng, Y. Deng, and H. W. J. Bl\"ote,
Phys. Rev. E {\bf 78}, 031136 (2008). 


\end{thebibliography}
\end{document}